\documentclass[reprint,pra,aps,10pt,longbibliography,showkeys]{revtex4-2}
\usepackage{natbib}
\usepackage{datetime}
\usepackage[pdftex]{graphicx}
\usepackage{xcolor}
\colorlet{blue}{black}
\colorlet{red}{black}
\def\be{\begin{eqnarray}}
\def\ee{\end{eqnarray}}

\def\({\left(}
\def\){\right)}
\begin{document}
\title{Shape fluctuations and radiation from thermally excited electronic states 
of boron clusters}
\author{T. H{\"o}ltzl}
\affiliation{Furukawa Electric Institute of Technology, 1158 Budapest, Hungary}
\affiliation{ELKH-BME Computation Driven Chemistry Research Group and 
Department of Inorganic and Analytical Chemistry, 
Budapest University of Technology and Economics, 1111 Budapest, Hungary}
\email{tibor.holtzl@furukawaelectric.com}
\author{P. Ferrari}
\affiliation{Quantum Solid-State Physics, Department of Physics and Astronomy, 
KU Leuven, 3001 Leuven, Belgium}
\author{E. Janssens}
\affiliation{Quantum Solid-State Physics, Department of Physics and Astronomy, 
KU Leuven, 3001 Leuven, Belgium}
\author{K. Hansen}
\affiliation{Lanzhou Center for Theoretical Physics, 
Key Laboratory of Theoretical Physics of Gansu Province, 
Lanzhou University, Lanzhou, Gansu 730000, China}
\affiliation{Center for Joint Quantum Studies and Department of Physics, 
School of Science, Tianjin University, 92 Weijin Road, Tianjin 300072, China}
\email{hansen@lzu.edu.cn,KlavsHansen@tju.edu.cn}
\homepage{http://www.klavshansen.cn/}
\date{\today,~\currenttime}
\begin{abstract}
The effect of thermal shape fluctuations on the recurrent fluorescence of 
boron cluster cations, B$_N^+$ ($N=9-14$), has been investigated numerically, 
with a special emphasis on B$_{13}^+$. 
For this cluster, the electronic structures of the ground state and the four lowest 
electronically excited states were calculated using time-dependent density functional 
theory, and sampled on molecular dynamics trajectories of the cluster calculated at an 
experimentally relevant excitation energy. 
The sampled optical transition matrix elements for B$_{13}^+$ allowed to construct 
its emission spectrum from the thermally populated electronically excited states. 
The spectrum was found to be broad, reaching down to 
{\color{blue}at least 0.85 eV}.
This contrasts strongly with the static picture, where the lowest electronic 
transition happens at 2.3 eV. 
The low-lying electronic excitations produce a strong increase in the rates of 
recurrent fluorescence, calculated to peak at $4.6 \times 10^{4}$ s$^{-1}$, with a 
time-average of $8 \times 10^{3}$ s$^{-1}$.
The average value is one order of magnitude higher than the static result, 
approaching the measured radiation rate. 
{\color{blue} Similar results were found} for the other cluster sizes. 
Furthermore, the radiationless crossing between the ground-state and the first 
electronic excited 
state surfaces of B$_{13}^+$ was calculated, found to be very fast compared 
to experimental time scales, justifying the thermal population assumption.
\end{abstract}
\maketitle

\section{Introduction}

Thermal radiation, long considered to be restricted to vibrational cooling, has 
also been demonstrated to occur from electronically excited states, via the so-called 
recurrent fluorescence (RF) or Poincare radiation. 
This type of radiation is attracting increasing attention as more and more molecules 
and clusters are recognized as emitters, {\color{blue} and in part also} 
stimulated by the implications for astrophysics. 
To date, recurrent fluorescence has been observed in fullerenes 
\cite{Hansen1996,andersen1996}, 
where the effect was first seen, as well as in polycyclic hydrocabon (PAH) molecules 
\cite{Martin2013,Bernard2019} and in clusters of several metallic and 
semiconductor elements. 
For very recent results of direct astrophysical relevance, please see
\cite{IidaMNRAS2022}.
Moreover, RF is known to be present in both anionic, neutral and cationic species, 
highlighting its general relevance at the nanoscale \cite{Ferrari2019}.

The existence of this type of radiation follows from time reversal once the presence 
of radiationless transitions is established. 
Radiationless transitions connect different Born-Oppenheimer surfaces, allowing 
for the phenomenon of internal conversion (IC), in which electronic energy is 
converted into vibrational motion on the electronic ground state. 
Time reversal requires that the reverse process, i.e., inverse internal 
conversion (IIC), is also allowed, which means that in principle any excited electronic 
state that undergoes 
IC can be reached non-radiatively from the electronic ground state \cite{Chernyy2016}. 
In RF, photon emission {\color{blue} occurs} from electronically excited states that are populated 
via IIC. 
As the radiationless transitions are very well established for a number of molecules,
the main concern about the presence of RF has been the question to which degree 
molecular electronically excited states are populated thermally.
The experimental evidence is now strongly in favor of the existence of RF, and the 
theoretical results here will add to this evidence.

Although RF was predicted several decades ago {\color{blue}
\cite{NitzanJCP1972,NitzanJCP1979,Leach1987}}, photons emitted {\color{blue} by} 
this mechanism from mass selected clusters and molecules were only 
{\color{blue} observed recently,} for the molecules C$_6^-$ \cite{Ebara2016}, C$_4^-$ 
\cite{Yoshida2017} and naphthalene {\color{blue} cations}\cite{Saito2020}. 
The vast majority of other studies of RF have deduced its presence from 
its quenching effect on the unimolecular decays of the emitting particles
{\color{blue} (see ref. \cite{Ferrari2019} for a review of this method)}.
The involved unimolecular {\color{blue} signature} 
decays can be the loss of atoms or larger fragments or, for anions and neutrals, 
thermal electron emission.
The observed time scales for RF vary strongly with the system, from the highest rate 
constants found so far of $10^6$ s$^{-1}$ for cationic gold 
\cite{Hansen2017} and cobalt \cite{Peeters2021} clusters, to close to the typical 
vibrational radiative time scales that are four orders of magnitude slower. 

Two of the three cases where RF photons {\color{blue} have been} detected directly 
from mass selected molecular beams {\color{blue} (C$_6^-$ and C$_4^-$) } 
employed a {\color{blue} narrow detection window centered on} the wavelengths 
where the particles in their ground-state absorb light \cite{Ebara2016,Yoshida2017}.
The selective detection of this narrow spectral region was motivated by the desire to 
establish the origin of the photons. 
In the experiment measuring the spectrum of photons emitted from naphthalene, a 
significant broadening relative to the ground state absorption spectrum  
was observed \cite{Saito2020}. 

Such broadening is not unexpected and can be understood qualitatively as the 
consequence of the varying geometric structures sampled by the thermal 
excitation, with the concomitant sampling of transition energies and oscillator 
strengths across the involved potential energy surfaces. 
Spectral smearing is thus expected to be intrinsic to the RF phenomenon.

The measurements on the quenching effect of radiation on competing processes 
usually only provide a single value in the form of the photon emission rate constant.
This value is the thermally averaged photon emission rate constant 
at the total excitation energy where it equals the unimolecular decay rate
constant, $k_{\rm ph}\sim k_{\rm unimol}$, because this defines the energy
where the emission rate constant is measured (see Ref. \cite{Ferrari2019} for
a detailed explanation of this).

Electronically excited states characterized spectroscopically tend to misrepresent 
emission rate constants, because these data usually only pertain to the very 
limited part of configuration space that correspond to excitation from a ground 
state geometry.
In contrast, highly excited and free-roaming species will explore much wider 
{\color{blue} regions of configuration space}.
The effective thermally-averaged transition energies and oscillator strengths will 
therefore in general differ from values derived from low temperature spectroscopic 
data.
Hence, the thermal exploration of potential energy surfaces provides information 
on their shape.
The investigation of this effect for well characterized systems is the main motivation 
of this work. 

The system chosen for {\color{blue} this} study is boron clusters, for which 
the effective (averaged) photon emission rate constants, but no emission spectra, 
have been measured upon laser excitation \cite{FerrariBoron2018}. 
The experimentally measured radiation rate constants for these clusters are 
significant, reaching for example $6 \times 10^{4}$ s$^{-1}$ for B$_{13}^+$ 
and $5 \times 10^{5}$ s$^{-1}$ for B$_{9}^+$. 
These high values strongly suggest that the emission mechanism is RF, and indicate
that the clusters possess optically active electronically excited states at low energies.
In this work, most attention is given to B$_{13}^+$, for which the ground-state 
geometry {\color{blue} has been determined based on spectroscopic 
experiments} \cite{Fagiani2017}. 
We expect that the general picture revealed for this cluster apply equally well to 
other boron clusters which, noted in passing, by themselves are very interesting 
objects, with remarkable planar geometries, possessing fascinating fluxionality 
properties \cite{Fagiani2017,Oger2007,Zhai2014,Kiran2009,Li2017}. 
Moreover, low-lying quasi-degenerate electronic states in boron clusters 
have also been shown to play an important role in their electron wavepacket 
dynamics \cite{Arasaki2019}.

\section{Thermal shape fluctuations and electronic structure}

\begin{figure}[htp]
\vspace{-0.5cm}
\includegraphics[scale=0.35]{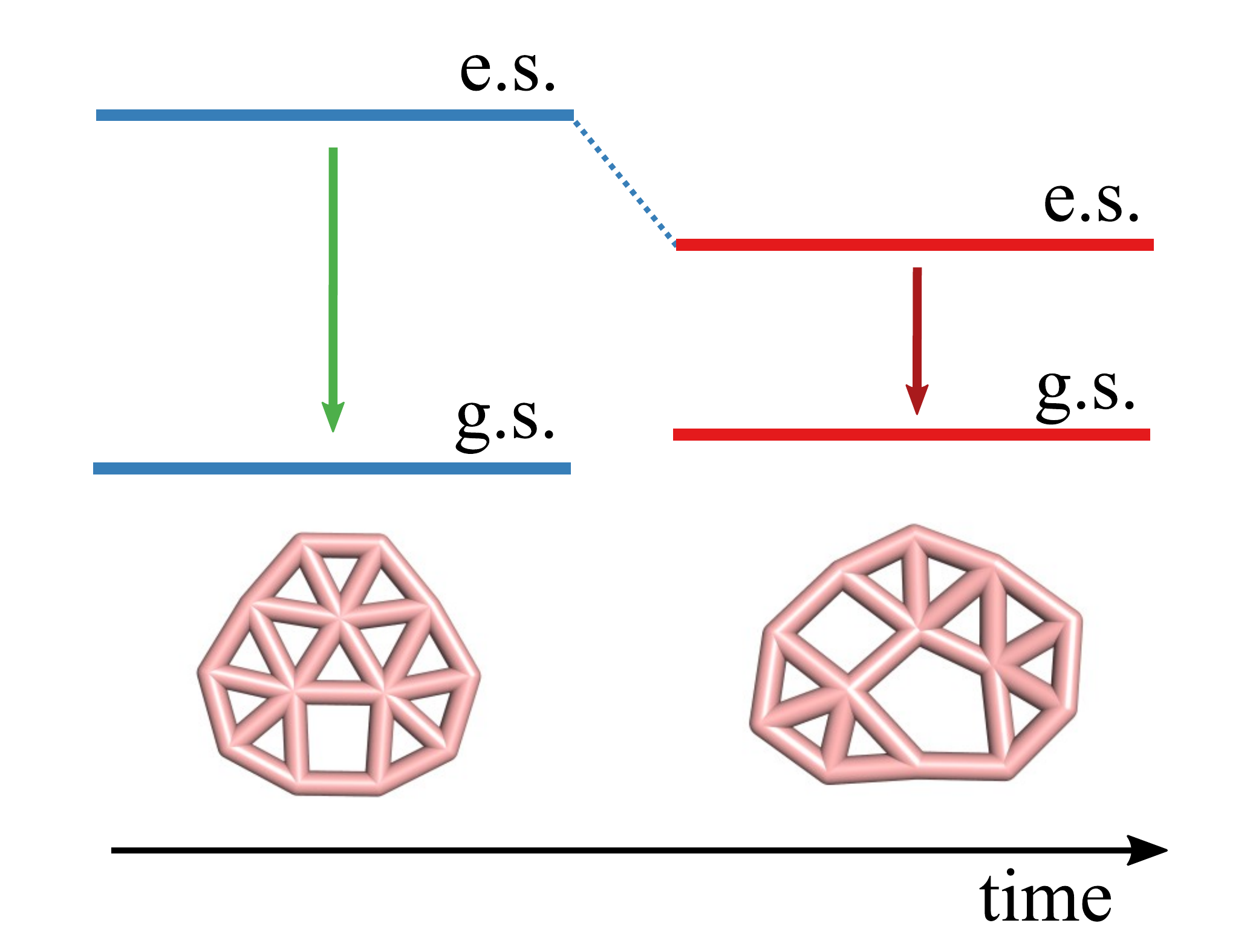}
\begin{centering}
\caption{Schematic illustration of the possible reduction of the energy difference 
between the ground state and the first excited state of a cluster, caused by 
temperature-induced geometrical deformations.
\label{Scheme}}
\end{centering}
\end{figure}

The central idea of the work is illustrated schematically in Fig. \ref{Scheme}. 
For a cluster in its ground-state geometry with a large energy difference between 
the electronic ground-state and the first electronic excited state (left part of the figure),
an optical transition between such states will translate into a small radiation rate 
\cite{Hansen2018book}.
This is not consistent with the observed high photon emission rate constants of, for 
example, laser-excited boron clusters \cite{FerrariBoron2018}.
However, the geometry of highly excited clusters will fluctuate over time as illustrated 
schematically on the right side of the figure.
This dynamics can reduce the energy difference between the ground state and the 
{\color{blue} electronically excited states}.
This will cause a higher thermal population of the excited states, which results in faster 
photon emission \cite{Ferrari2019}.

An explicit calculation of this reduction of the excitation gap (electronic transitions 
modified by structural deformation) has already been performed for 
Al$_{13}^-$, based on a spherical box potential of finite depth, undergoing a 
spheroidal distortion \cite{Kresin2006}, albeit without any considerations of thermal 
radiation.
The work showed a reduction in the HOMO-LUMO gap energy caused by this 
structural deformation.

\section{Computational details}

To investigate this thermal deformation-induced effect in boron clusters, we have 
performed {\color{blue} Molecular Dynamics (MD) simulations with 
potential energies calculated with {\color{red}Density Functional Theory} 
(DFT).
The MD simulations are standard integration of the Newtonian equations of motion.}
Specifically, the linear-response time-dependent DFT (LR-TDDFT) formalism was 
employed in order to compute the lowest lying low spin excitations in B$_N^+$ 
($N=9-14$) clusters. 
For B$_{13}^+$ four electronic excitations were calculated. 
Due to the long computer times required for these calculations, however, only 
two electronic excitations were computed for the other cluster sizes.
The computations were performed using the CAM-B3LYP exchange-correlation 
functional \cite{Yanai2004}, in combination with the def2-SVP basis set 
\cite{Weigend2005}. 
The relatively small basis set allows for {\color{blue} relatively} long MD 
simulation times. 
CAM-B3LYP was found to give excellent excitation energies for boron clusters, 
based on a benchmarking with coupled cluster theory (EOM-CCSD) \cite{Shinde2016}, 
while our computations showed only a small difference in the excitation spectra 
between the split valence def2-SVP and the triple zeta valence def2-TZVP basis sets. 
For example, in B$_{13}^+$, the energy of the first {\color{blue} vertical} 
electronic excitation {\color{blue}from the lowest energy singlet geometry}
 (S$_0$ $\rightarrow$ S$_1$) is calculated to be 2.27 eV and 2.25 eV, for def2-SVP and 
def2-TZVP, respectively. 

The ground state geometry of B$_{13}^+$ was experimentally identified in 
Ref. \cite{Fagiani2017}. 
{\color{blue} The molecular dynamics simulations on the first excited low spin 
potential energy surface were performed with a microcanonical ($NVE$) 
ensemble, with
initial coordinates of the lowest energy structure on the ground state potential energy 
surface (vertical excited state) and with initial nuclear velocities selected from a 
canonical energy distribution. 

The choice of using a temperature for a microcanonical simulation was made 
because the decay dynamics of the beam defines an effective, microcanonical 
temperature, and this temperature can be determined.
The relation was suggested by Gspann \citep{Gspann} and developed  
by Klots \citep{klotsNature1987,KlotsZPD1991}. 
It reads (with $k_{\rm B} =1$):} 
\be
\label{temp}
T=D(1/\ln(\omega_{\rm a}t) +1/2C_{\rm v}).
\ee
{\color{blue} The parameters in this expression are the evaporative activation energy, $D$, 
the frequency factor of the unimolecular rate constant, $\omega_{\rm a}$, the experimental 
measurement time $t$, and the cluster's heat capacity, $C_{\rm v}$, given 
in units of Boltzmann's constant. 
In these experiments the time $t=1/k_{\rm p}$ is the value at which the thermal photon 
emission rate constant is equal to the unimolecular fragmentation rate constant,
and the time is therefore set to this value. 
In brief, at higher excitation energies, instantaneous fragmentation is the dominant 
cooling channel, whereas at lower values, fragmentation is exponentially suppressed.
Hence
\be
\label{temp2}
T=D(1/\ln(\omega_{\rm a}/k_{\rm p}) +1/2C_{\rm v}).
\ee
and this is therefore the energy at which the photon emission rate constant is 
determined experimentally.}

{\color{blue}For completeness we should mention that this temperature is 
the microcanonical value, but the difference to the canonical temperature 
can be ignored here.
For more details on this the reader can consult \citep{Hansen2018book}, 
and ref. \citep{AndersenJCP2991} for the meaning of the microcanonical temperature.}

{\color{blue}The value found for B$_{13}^+$ is 3855 K.
In energy units this temperature is 0.33 eV.
This temperature together with the speeds selected at random from the Maxwell-Boltzmann
distribution gave the total {\color{red} nuclear kinetic energy} of 5.61 eV, close to the canonical average 
value of 5.48 eV.
The simulations were started from the ground state geometry and the first excited state.
The energy of that state, 2.27 eV, should be added to the total energy. 

The potential energy surface of B$_{13}^+$ and the possible emission routes 
were explored using adiabatic molecular dynamics simulations on the S$_1$ state. 
The molecular dynamics (MD) simulations propagate the system in time 
by solving the classical equations of motion for the nuclei,
while calculating the potential energy on all surfaces point-by-point with the 
density functional theory (DFT) calculations mentioned above.
The S$_{\rm 1}$ surface was chosen as a representative of an emitting surface,
but the specific choice is not critical for the exploration of the phase space.}
The simulations were tested with two different time-steps, 0.02 and 0.1 fs. 
The results were very similar, and longer simulations were therefore 
conducted with the latter value.

The small S$_0$-S$_1$ energy differences that are calculated (see Section III) 
imply that non-adiabaticity is important.
An accurate description of such effects {\color{blue} is computationally 
demanding, and non-adiabatic effects are therefore} explored only for B$_{13}^+$. 
To calculate these effects, and ensure that the computations are reliable, 
three different methods were used, namely spin-flip Density Functional Theory 
(SF-DFT) employing the recommended Becke half \& half LYP (SF-BH\&HLYP) 
functional and the def2-SVP basis, the spin-flip equation of motion coupled clusters 
singles and doubles (SF-EOM-CCSD) \cite{Casanova2020} and the {\color{red} def2-SVP} 
basis set, and the extended multi-state second order complete active space 
(XMS-CASPT2) method \cite{Shiozaki2011}, based on state-averaged CASSCF 
reference involving the lowest three {\color{blue} singlet} 
states and the TZVP basis set.
It has been shown recently that XMS-CASPT2 and SF-BH \& HLYP yield similar 
results {\color{blue} for conical intersections} \cite{Winslow2020}. 
The DFT, SF-DFT and SF-EOM-CCSD computations were performed using the 
Q/Chem 5.2 and 5.4 program packages \cite{Epifanovsky2021}, while the 
XMS-CASPT2 computations were completed using the BAGEL code 
\cite{Shiozaki2018}{\color{blue}, with results available in the Supporting 
Information (Table S1).

These computations make it possible to estimate the transition rate through 
{\color{blue} the} minimum energy crossing {\color{blue} points} 
{\color{red} toward  the S$_0$ and the  S$_1$ state}.
The magnitude of these time constants indicates that on the much
longer microsecond experimental time scale the excited states can {\color{blue} 
reach thermal populations.}
This allows us to {\color{blue} sample the phase space using molecular dynamics 
simulations and estimate the radiative emission rate from the thermal populations
of the excited states}}

Nonadiabatic couplings (NACs) can be computed by differentiating the electronic 
wavefunctions with respect to the nuclear coordinates. 
Here we used the pseudo-wavefunction approach to compute the NACs analogously 
\cite{OuJCP2014,ZhangJCP2014}, using TDDFT employing the SF-BH\&HLYP 
method, as implemented in the Q-Chem software \cite{Epifanovsky2021}.

\section{Dynamics of electronic excited states}

\begin{figure}[htp]
\includegraphics[scale=0.35]{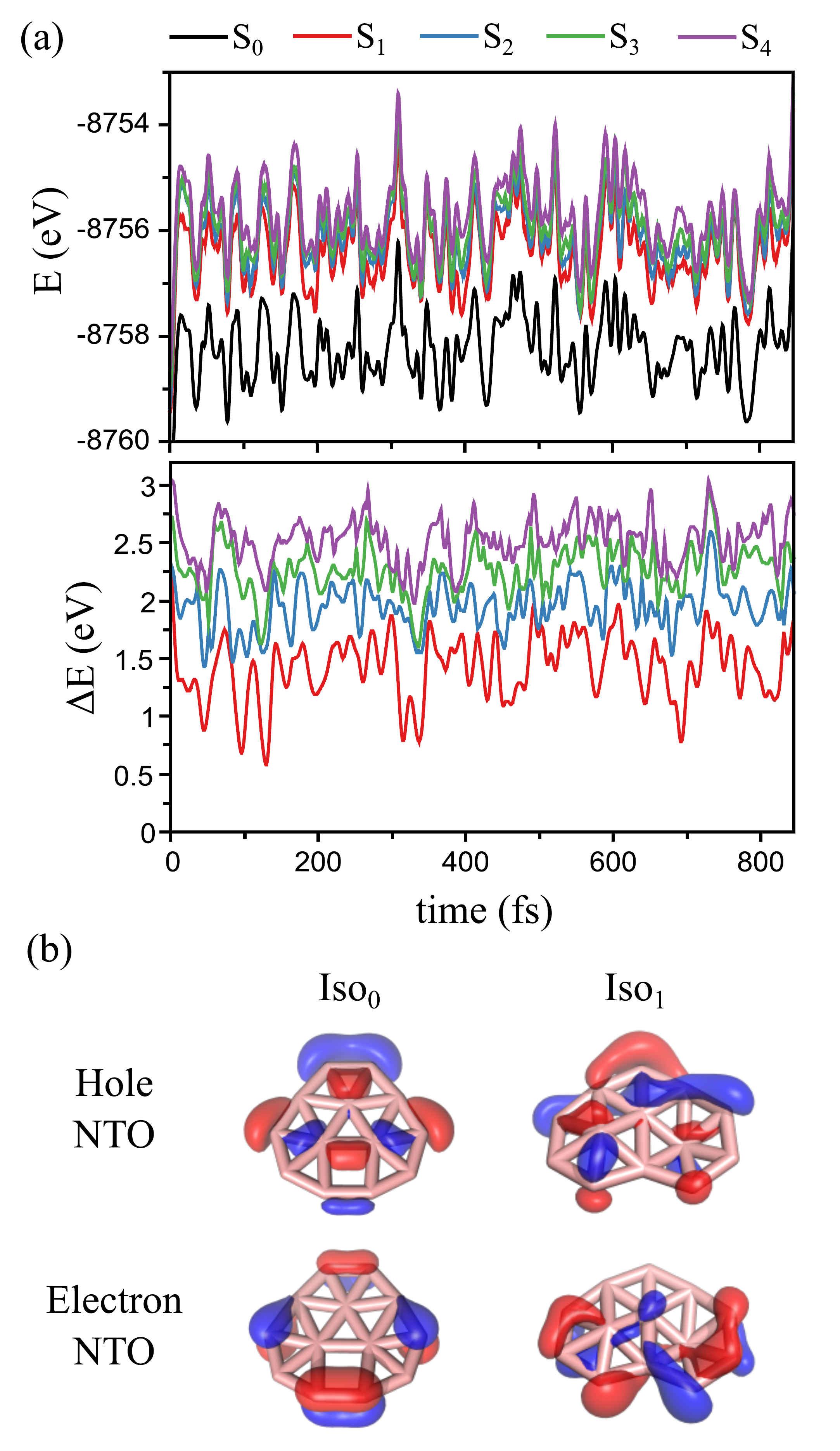}
\begin{centering}
\caption{(a) Top: Energies of the electronic excited states S$_i$ ($i=1-4$) 
and the ground state S$_0$ of B$_{13}^+$, as a function of simulation 
time. 
Bottom: Energy difference between each electronic excited states and the ground 
state. 
(b) Cluster geometries and the isosurfaces (isovalues: 0.05) of the leading 
Natural Transition Orbital (NTO) pairs for the S$_0$ ground state at the lowest 
energy geometry (initial structure, iso$_0$ at $t=0$) and at a low energy structural 
isomer (iso1 at an optimized intermediate structure, c.f. Fig. 3).\label{MDsimulation}}
\end{centering}
\end{figure}

The top panel of Figure \ref{MDsimulation}a shows the energies of the four 
lowest electronic states, in addition to the S$_0$ ground-state, along the 
molecular dynamics trajectory for B$_{13}^+$. 
The bottom panel presents the energies of the excited electronic states relative to 
the energy of S$_0$. 
Interestingly, the four excited states and the ground-state show a correlated pattern 
with time. 
{\color{blue}As seen from the figure, the energy difference between the first 
excited state and the ground state (red curve) at the initial configuration, and 
thus the lowest vertical excitation energy in the ground state geometry, 
is over 2 eV.
This} energy difference reduces to below 1 eV at several later moments, 
showing the possibility of emission of lower energy photons.
Whether this possibility is realized depends on whether the states are 
thermally populated, which in turn depends on the excited state energy relative to 
the ground state energy and not on their difference with respect to the simultaneous
energy of S$_0$.

Geometry optimization starting from the regions with {\color{blue} a}
small S$_0$-S$_1$ energy 
difference confirmed the existence of a structural isomer with an energy which is only 
slightly higher than the lowest energy isomer, as shown in Figure \ref{MDsimulation}b. 
In both cases, the excitation from the ground to the first excited state can be 
described approximately by a single electron configuration, which involves the 
formal electron transfer from the highest occupied molecular orbital (HOMO) to 
the lowest unoccupied molecular orbital (LUMO). 
This is accurately reflected by the fact that the excitation can be well described 
using a single natural transition orbital (NTO) pair, where the shape of the hole 
and the electron NTOs resemble those of the HOMO and LUMO, respectively. 
It is clearly visible that the shapes of the electron and hole NTOs follow the 
distortion of the cluster shapes towards the low energy structural isomer. 

The analysis shown so far involves the adiabatic approximation i.e. the MD 
trajectory was propagated on a single potential energy surface, with the different 
surfaces computed independently. 
However, it has been shown that the dynamics of boron clusters also involves 
non-adiabatic effects \cite{Arasaki2019}, including the presence of several conical 
intersections and avoided crossings.
These were therefore investigated in more detail.

\section{Non-adiabatic effects}
\label{non-adiabaticeffects}

{\color{blue} The relevance of excited states in the description of the radiative cooling
of a cluster or molecule is to some degree a question of time scales.
States need to be populated. 
To assess whether excited states are populated, the time scale determined by the 
coupling between the should be compared with the microsecond or longer experimental 
time scale at which the populations are monitored, i.e. the time at which the radiative 
cooling is measured. 
This long time scale puts very mild conditions on the coupling.

In the current section we explore a part of the potential energy surface, at which the 
relative populations of the Born-Oppenheimer (BO) states are calculated 
by calculating non-adiabatic couplings.
As the typical time scale of RF cannot be attained using conventional molecular 
dynamics simulations, we explore the stationary {\color{blue} (zero derivative) points
and the}
crossing points between the lowest two singlet states in the neighborhood of the lowest 
energy geometry. 
The crossing points on the computed trajectories (described in the caption of Figure 3) 
clearly show the possibility of non-radiative transitions between the two lowest
singlet states. 
Crossings between higher excited states should also present, but these do not alter 
the results found here for the two lowest states fundamentally.}

\begin{figure}[htp]
\vspace{-0.5cm}
\includegraphics[scale=0.35]{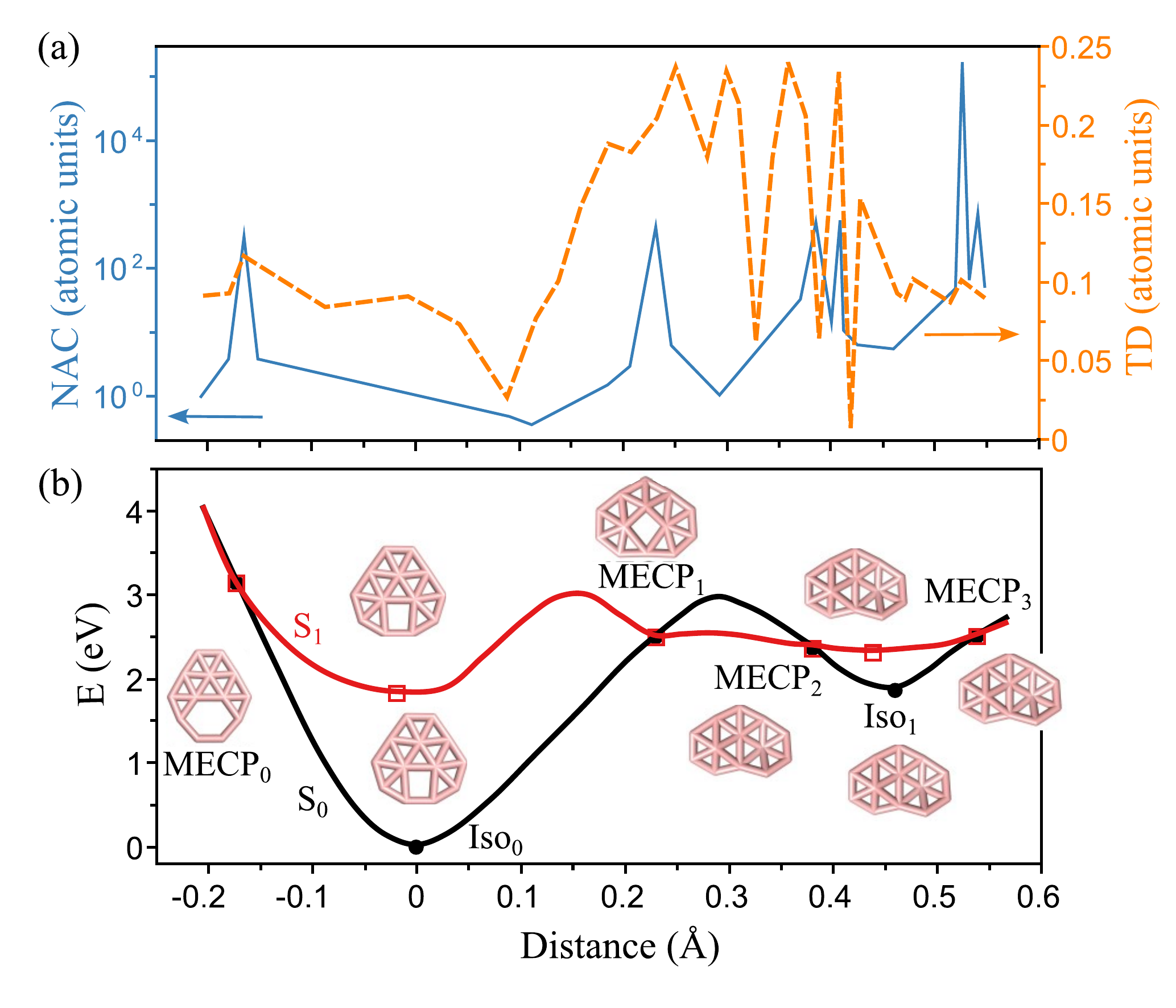}
\begin{centering}
\caption{(a) Magnitude of the transition dipole moments (TD) and the non-adiabatic 
coupling vectors (NAC, logarithmic scale), both in atomic units 
{\color{blue} (Hartree/$a_0$ for NAC) along the 
constructed linear synchronous interpolation curve described in the main text.
The abscissa is the root-mean-square of the stretched nuclear coordinates}. 
(b) Geometries and energies along {\color{blue} the path} between the 
lowest energy Iso$_0$ and the intermediate Iso$_1$ geometries, and the 
adjacent {\color{red} minimum energy} crossing points of the two lowest potential surfaces 
(S$_0$ and S$_1$). 
Stationary and {\color{red} minimum energy crossing points are indicated} by dots. 
The paths from {\color{blue} the isomer} Iso$_0$ to MECP$_0$ and from Iso$_1$ 
to MECP$_3$ are stretched for visibility. 
Calculations are performed at the SF-BH\&HLYP/def2{\color{blue}-SVP level.
The S$_0$-S$_1$ excitation energy at zero abscissa differs from the previously quoted 
0.225 eV and 0.227 eV because the value here is not vertical} {\color{red} i.e. 
the S$_0$ and S$_1$ geometries are different, except in the case of the MECPs.}
\label{Crossing}}
\end{centering}
\end{figure}

The small energy difference between the S$_0$ and the S$_1$ states of 
B$_{13}^+$ {\color{blue} will} induce large non-adiabatic couplings, and also 
generate conical intersections. 
Here, we have optimized the initial and the isomeric geometries of 
B$_{13}^+$, and systematically explored the {\color{blue} two} 
adjacent minimum energy crossing points (MECP, Figure \ref{Crossing}). 
{\color{blue} 
The MECPs are the lowest energy points on the line defining the intersection of the 
two surfaces.  
The variations in potential energies are much below the energy used in the MD simulations. 
Hence the crossing configurations shown are well within energetic reach.
The computations were performed using SF-TDDFT, which was 
motivated by the fact that unlike the conventional TDDFT, this method correctly 
describes the topology of the conical intersections involving the S$_0$ state 
\cite{LevineMP2007}.}
{\color{red}To further assess the accuracy of our computations, we recomputed the 
geometries and energies of the minima and the MECPs} using SF-EOM-CCSD and 
XMS-CASPT2 methods (see Table S1 in the Supporting Information).

{\color{red} To illustrate the topology of the S$_0$ and the S$_1$ potential 
energy surfaces we computed
the linar synchronous transit paths (i.e. linear interpolation in Cartesian coordinates)
between the minima and the MECPs shown in Fig. \ref{MDsimulation}(b).}

{\color{blue}The abscissa in both curves is the root mean square deviation of the atomic 
positions compared to Iso$_0$.
Note that this figure is not a trace of the MD simulation, and that the data do not
enter the statistics derived from that simulation.}
Figure \ref{Crossing} shows that there are several MECPs between the S$_0$ 
and the S$_1$ potential energy surfaces on this curve. 
The energies of those two states at the isomeric structures are not much higher 
than the ground state energy in the lowest energy minimum, and {\color{blue} 
will thus be easily} accessible at the experimental conditions of radiative cooling 
studies of excited clusters in molecular beams, {\color{blue} as discussed above.
This is confirmed by calculations of the rate coefficients for crossing, involving thermally 
activated reactions (see Supporting Information for details).}

The magnitude of the transition dipole moment between the S$_0$ and the S$_1$ 
states determines the rate of photon emission via {\color{blue} the electronic transition
between the two states}, while the magnitude of the non-adiabatic coupling indicates 
the efficiency of the non-radiative relaxation and the excitation. 
It is clearly seen from Figure \ref{Crossing} that, as expected, the probability 
of the non-radiative relaxation {\color{blue} or excitation} is very high in the 
vicinity of the MECPs, as the magnitude of the non-adiabatic couplings increases 
by factors up of  $\sim$ $10^2$ to $10^4$ compared to the original value, while 
the magnitude of the transition dipole moment {\color{blue} increases by a more modest
factor of} 2-3. 
Hence, although photon emission is more probable near {\color{blue} MECPs and the
conical intersections that may be present} than elsewhere, the non-radiative 
{\color{blue} coupling} is much more enhanced at these points. 
In this connection it is worth noting that non-adiabatic wavefunction dynamics of boron 
clusters have indicated the presence of several conical intersections and frequently 
occurring high non-adiabatic couplings \cite{Arasaki2019}.
{\color{blue} Transition state theory calculations (see the Supporting Information for 
more details)  show that the transition rate at the MECP$_1$ is approximately 
$4 \times 10^9$ s$^{-1}$  and thus well within the microsecond time scale of the 
emission.}
{\color{blue} These results showing a rapid exchange of energy between the 
electronic ground-state and the first-excited state of B$_{13}^+$ strongly support
the suggestion that photon emission in boron clusters proceeds via IIC and recurrent 
fluorescence} in boron cluster cations.

\section{Microcanonical averaging}

The combination of time-dependent energies of the electronic {\color{blue} state 
S$_0$ ($i=0-4$) energies} with the oscillator strengths $f_i$ 
for the transitions, allow the construction of the emission spectrum of B$_{13}^+$
{\color{blue} for the sampled geometries}.
In the rapid equilibration scheme {\color{blue} relevant here} the statistical weight of 
an excited state is given by the level density of the kinetic energy for that geometry.
With $E$ the total (conserved) excitation energy relative to the absolute ground state, 
and $V_i(x)$ the potential energy of the state $i$ {\color{blue} ($i=0-4$) in the geometry
labeled by $x$},
the unnormalized populations of the states ($g_i$) are, with $s\equiv 3N-6$, equal to
\be
g_i(x) = c(E-V_i(x))^{s/2-1}, i=0,1,....
\ee
{\color{blue} This expression is simply the density of states 
for $s$ momenta with a total energy $E-V_i(x)$ and the sampled geometry
in the simulated microcanonical ensemble.
The assumption is the fundamental postulate of thermodynamics that all quantum 
states are populated with equal probability.}
The constant $c$ is identical for the different states, as are the number of degrees 
of freedom of the nuclear motion, $s$. 
The population of the state $i$ is therefore
\be
p_i = \frac{g_i(x)}{\sum_{j=0} g_j(x) }, 
\ee
where $j$ runs over all relevant states. 
For values of $V_i(x)$ exceeding $E$, obviously state $i$ is unpopulated.

{\color{blue} With the oscillator strengths and energies of the surfaces the rates of 
photon emission can thus} be computed as a function of time {\color{blue} along the trajectory. 
At each point t}he photon emission rate constant from state $i$ {\color{blue} ($i-1-4$)}
 is given by \cite{Hansen2018book}:
\be
k_{\rm p}^i = 7.421 \times 10^{-22}{\rm Hz} f\nu_{0}^2 p_i
\ee
In this expression, $f$ is the oscillator strength and $h\nu_0$ the transition energy. 
For B$_{13}^+$ the total rate constant is then calculated as the sum 
of the four $k_{\rm p}^i$ values {\color{blue}(see 
Figure \ref{MDsimulation}a)}. 
For the other {\color{blue} cluster sizes two electronic excited 
states were computed.}

\section{Emission spectra}

\begin{figure}[htp]
\vspace{-0.2cm}
\includegraphics[scale=0.35]{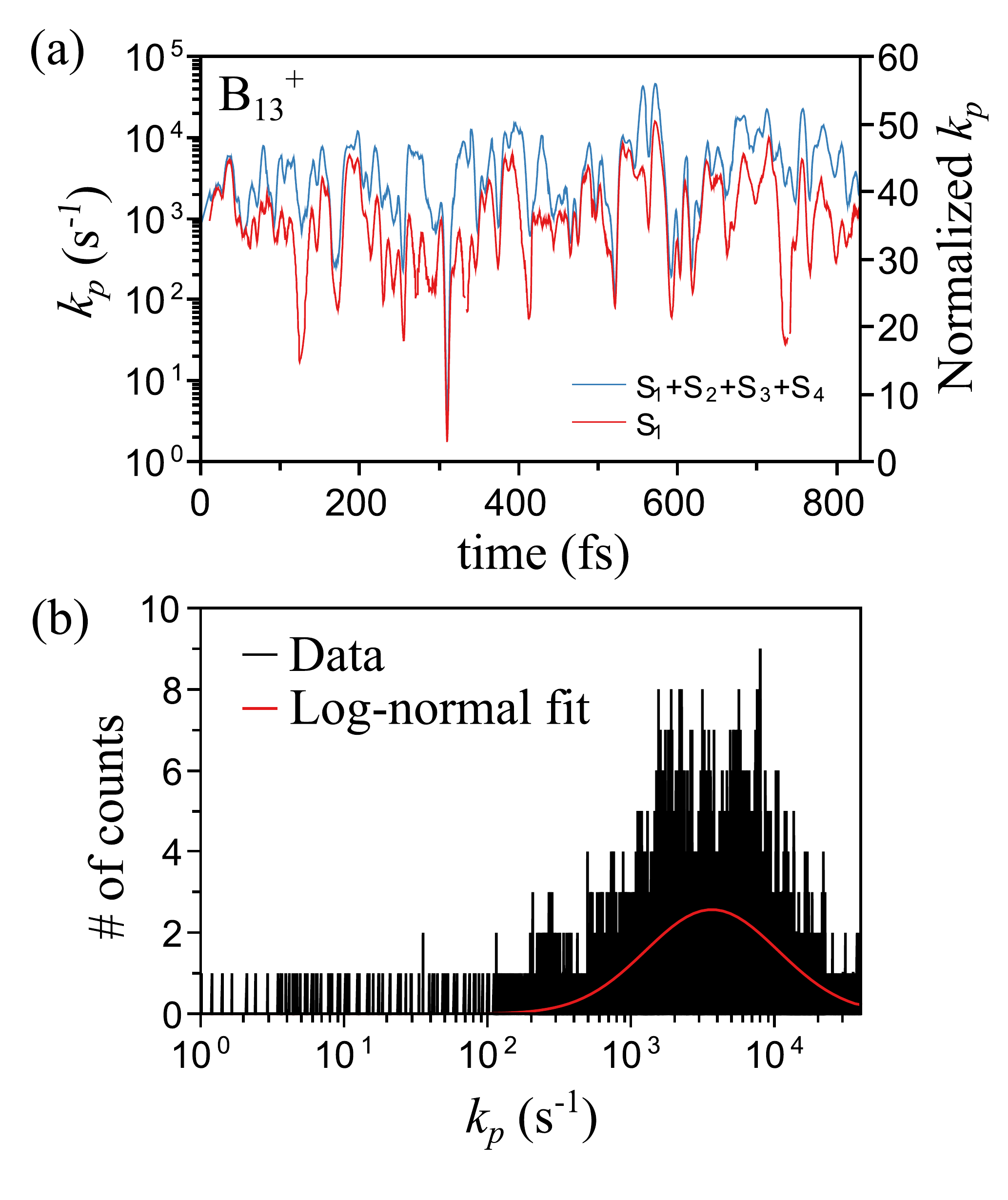}
\begin{centering}
\caption{(a) Time-dependent radiation rate constant calculated for B$_{13}^+$ at an 
initial {\color{blue} kinetic energy corresponding to the} temperature of 3855 K.
{\color{blue} The red curve shows the radiation constant when only emission from 
S${\rm 1}$ is considered. 
The blue gives the emission rate constant for the first four excited states.}
(b) Distribution of $k_{\rm p}$ for B$_{13}^+$ as derived from the time-dependent 
behaviour in (a). 
A log-normal distribution is {\color{blue} used to fit} an average $k_{\rm p}$ value.  
\label{Rates}}
\end{centering}
\end{figure}

The result for B$_{13}^+$ is presented in Figure \ref{Rates}a),  where the total 
radiation rate constant is shown as a function of {\color{blue} MD simulation} time. 
{\color{blue}The figure also shows the effect of the number of excited states 
included in the calcuation.}
The {\color{blue} values vary} strongly with time, as expected {\color{blue} from}
the pronounced dependence of the energy of the excited states seen in 
Fig. \ref{MDsimulation}. 
The rate constant reaches significantly higher values than for the static picture in 
the ground-state geometry, as highlighted by the scale on the right hand axis of the 
figure, where $k_{\rm p}$ is normalized with respect to the rate at zero kelvin. 

Another important observation from this result is the quite high absolute values that are 
reached along the simulation (axis on the left), peaking at $4.6 \times 10^{4}$ s$^{-1}$. 
This value is very close to the experimental result in Ref. \citep{FerrariBoron2018}, 
which is $(6.1 \pm 2.8) \times 10^4$ s$^{-1}$. 
In Figure \ref{Rates}b, the distribution of $k_{\rm p}$ values is depicted 
and fitted by a log-normal distribution. 
This allows an estimation of the average $k_{\rm p}$, yielding the value
$8 \times 10^{3}$ s$^{-1}$.
{\color{blue} As discussed below, this improved estimate is still most probably an
underestimate of the values that are generated by the RF mechanism.}

\begin{figure}[htp]
\includegraphics[scale=0.35]{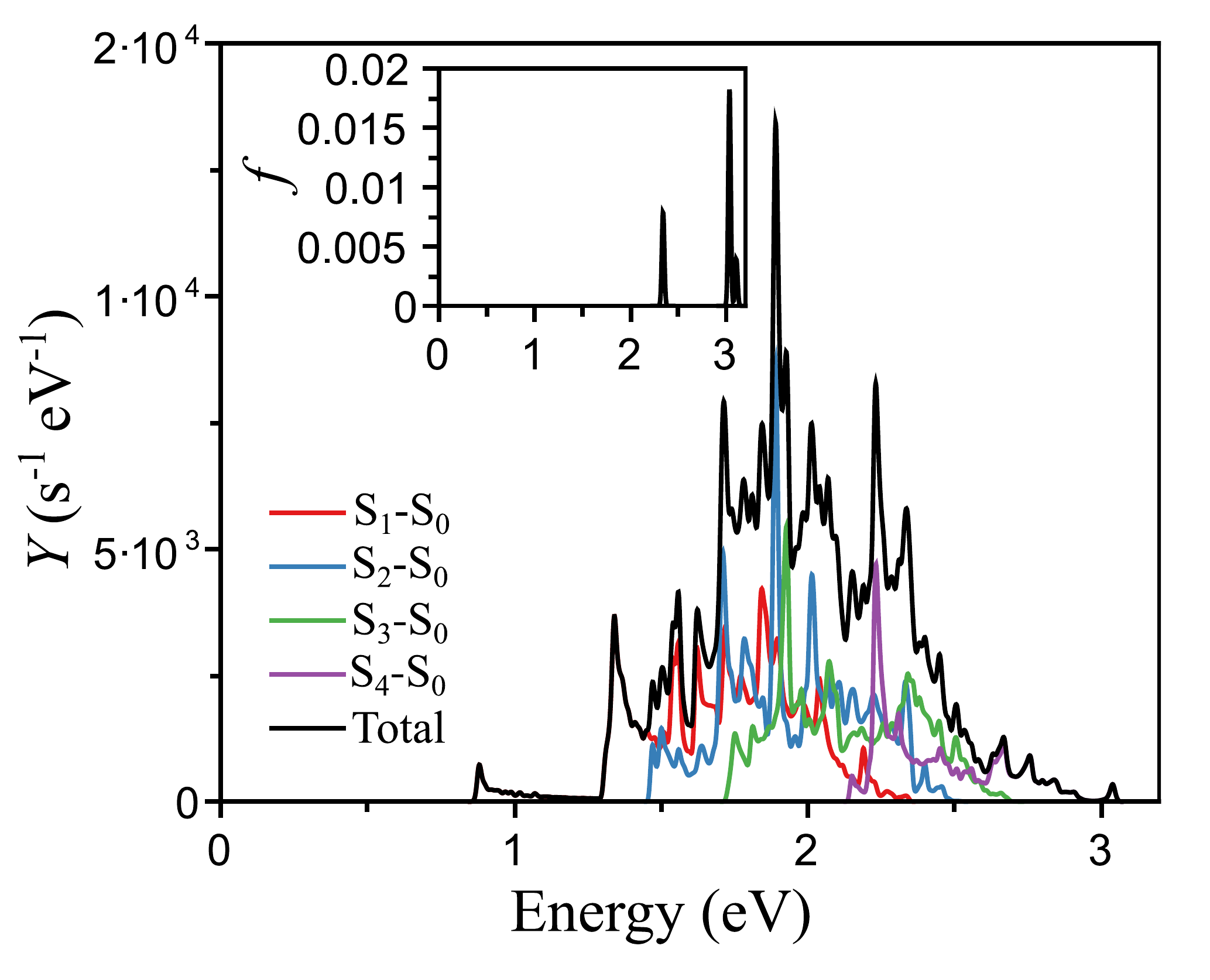}
\begin{centering}
\caption{Emission spectrum of B$_{13}^+$ at the microcanonical temperature of 
3855 K.{\color{blue} The spectrum is calculated with the difference of energy of the 
electronic states to the absolute ground state, yielding the populations of the 
excited states, and the oscillator strengths of those transitions.}
The sum $\sum_i Y_i$ is shown as a black line. 
The static emission spectrum, calculated with LR-TDDFT on the ground-state 
geometry of the cluster, is shown in the inset. 
This is constructed by assuming Gaussian functions around each transition, with 
an artificial width of 0.05 eV. \label{Spectrum}}
\end{centering}
\end{figure}

Based on the energies and oscillator strengths calculated for the four 
electronic states of B$_{13}^+$, an emission spectrum is constructed, as 
presented in Figure \ref{Spectrum}.
The spectrum is calculated as the sum over the spectra produced by emission from 
the four excited states {\color{blue} with the expression for each of them equal to}
\be
\label{i-spectra}
Y_i(h\nu) =\frac{\sum_j' k_{i,j} }{\Delta h\nu N}, 
\ee
where $N$ is the total number of sampled points and the primed sum over $j$ 
is over the points where the photon energy is found in the interval between
$h\nu$ and $h\nu + \Delta h\nu$.
{\color{blue} The total spectrum is then
\be
Y(h\nu) = \sum_i Y_i(h\nu).
\ee}
It shows bands of significant widths {\color{blue} for each of the excited states}, 
spanning {\color{blue}between 0.85 eV and 3 eV.
Note that the lower energy of 0.85 eV is the value observed in the relatively short 
simulations.
The true lower limit is most likely much lower, approaching close to zero
(see Fig. \ref{Crossing}).} 

This high-temperature spectrum is very different from the static picture calculated 
with LR-TDDFT at the ground-state geometry (shown in the inset). 
{\color{blue} That spectrum is composed of a small} set of discrete transitions, with the 
first occurring at 2.27 eV.
{\color{blue} It is worth reiterating that the spectrum in Figure \ref{Spectrum} represents 
the configurations encountered in the MD simulation.
The true lower energy cutoff will be lower, as demonstrated by the level crossings
seen in Fig. \ref{Crossing}.
The time-average, energy integrated total oscillator strength at the elevated 
temperature shown in Fig.\ref{Spectrum} is lower than the static values, but not significantly so.}

\begin{figure}[htp]
\vspace{0.0cm}
\includegraphics[scale=0.35]{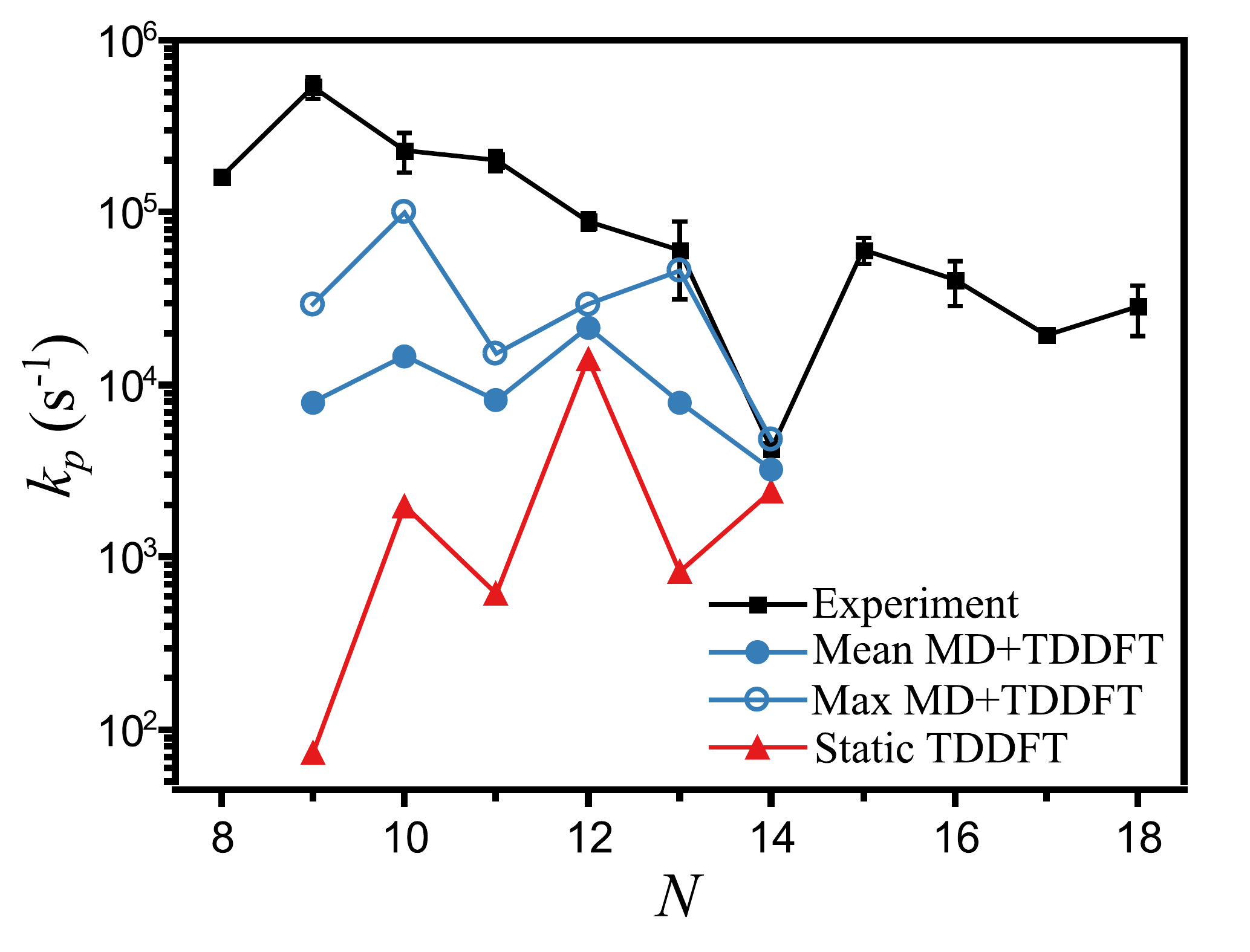}
\begin{centering}
\vspace{-0.5cm}
\caption{Radiation rate constants of B$_{N}^+$ clusters 
calculated with the MD+TDDFT analysis (blue circles). 
Filled and open circles correspond to the mean and the maximum values, 
respectively.  
For comparison, the experimental data from Ref. \citep{FerrariBoron2018} 
is reproduced in the figure (black squares), together with a calculation using 
TDDFT on the ground-state geometries (red triangles).
\label{BoronRates}}
\end{centering}
\end{figure}

The analysis performed for B$_{13}^+$ was also conducted for the other boron clusters 
{\color{blue} of the sizes $N=9-12,14$}. 
The results for {\color{blue} these clusters are included in the summary of all 
the radiative rate constants shown in Figure \ref{BoronRates},.
The figure compares the MD time-average rates obtained from the computations in 
this work with the experimental data of} Ref. \citep{FerrariBoron2018}. 
Additionally, the maximal values are presented as well as the rates calculated 
using LR-TDDFT on the (static) ground-state geometries. 
These are obtained from Ref. \cite{Oger2007}, and were re-optimized at the 
CAM-B3LYP/def2-SVP level. 

Although not all the size-dependent trends of the experimental data are reproduced by 
the MD+TDDFT simulations, the improvement relative to the static geometries is 
significant.
The improvements for the sizes 9, 10, 11 and 13, in particular, are considerable. 
For B$_{13}^+$, all four states contribute with a similar magnitude to the total emission 
rate constant.
We therefore expect that the restriction to {\color{blue} a limited number of excited states}
is responsible for part of the remaining discrepancy {\color{blue} between experiment and theory},
in addition to the role of non-adiabatic effects which were not included in this calculation
but known to be relevant, as discussed in section \ref{non-adiabaticeffects}.
{\color{blue} These are, briefly, the uncertatinty of the frequencies (described briefly in 
the SI) and the accuracy of the computation (SF-DFT and the more accurate XMS-CASPT2).
In addition, the branching ratio onto the two curves at the MECP is associated with 
large uncertainties.}

We can {\color{blue} also not} exclude that a longer MD run will also contribute higher intensity 
parts of the phase space, given the approximate log-normal shape of the $k_{\rm p}$ 
distribution. {\color{blue} This suggests} a long tail toward high values.

{\color{blue}
\section{Summary and perspectives}}

In this work, the role of temperature on the radiative cooling of boron clusters was 
investigated by means of molecular dynamics simulations in combination with 
time-dependent density functional theory and high-level ab-initio computations. 
With a focus on the computational analysis of the B$_{13}^+$ cluster, it was 
shown that at the effective temperature of 3855 K, estimated from experimental 
parameters, geometry fluctuations {\color{blue} cause} significant changes in the electronic 
structure of the cluster. 
At several points in the explored geometry, the energy of the low-lying electronic 
excited states is reduced by a significant amount, with a concomitant exponential 
increase in the recurrent fluorescence radiation rate.
Moreover, the possibility that energy is transferred non-radiatively via IC and IIC 
was explicitly demonstrated, by large non-adiabatic coupling vectors when the 
electronic ground and the first electronic excited state cross.

The thermal effects have two important consequences.
One is that the rates estimated in this work reduce the discrepancy between theory and 
experiment significantly, giving a strong indication of the importance of the effect and 
the modification of the RF values relative to ground state properties.

The second consequence is {\color{blue} the occurrence of a significant broadening 
and} redshift of the emission spectrum relative to that of the ground state.
The redshift seen in the thermalized spectrum must also be expected to be a 
general occurrence.
The lowest energy geometries tend to be those that maximize the HOMO-LUMO gap for 
molecules and clusters, and thermal excitation tend to reduce this gap, and 
hence generally reduce the excited state energies.
The tendency is amplified {\color{blue} in the emission spectra by the fact that}
low energy states tend to be more populated by simple phase space arguments. 
Hence, also this effect will favor a redshift of the spectrum.

There is therefore all reason to believe that the thermal effects described 
here on the emission rate constants and on the spectral shapes 
{\color{blue} will both} be present 
in other systems that dissipate energy by recurrent fluorescence.
It should also be noted that the deformations {\color{blue} explored in 
Fig. \ref{Crossing}} imply energy of emitted 
photons reaching down to {\color{blue} close to} zero due to level crossings.
{\color{blue} Measured spectra will therefore show the effect even more strongly 
than calculated here.

{\color{blue}
The internal excitation energy was determined as the sum of kinetic energies sampled 
from the canonical ensemble with the added energy of the S$_{\rm 1}$ state in the 
ground state geometry. 
The choice of using the S$_{\rm 1}$ state and not the electronic ground state as the 
initial configuration compensates for the fact that a fraction of the clusters are in 
electronically excited states, both S$_{\rm 1}$ and the higher states, and that this 
excitation energy should be added to the nuclear kinetic energy of 5.61 eV.

It also compensates for the fact that the deformed clusters, that take up most of the 
phase space, will have nuclear potential energies above those of the harmonic motion 
of the ground state.
It is not possible to know the precise value of this contribution before the entire phase 
space is sampled or the caloric curve determined experimentally, but an estimate can be 
obtained by considering the melting enthalpy.
The bulk melting point is 2349 K, which is below the temperature here, suggesting 
that the clusters are liquid, or in the equivalent finite sizes particle phase.
The melting enthalpy for bulk is 0.52 eV per atom for bulk.
Adding this would more than double the excitation energy.
The thermal estimates calculated here are therefore of a very conservative nature, 
which should be kept in mind when comparing them to the experimental values.}
 
Measurements of recurrent fluorescence spectra is a field in its infancy, with 
only one spectrum reported so far {\color{blue}\cite{Saito2020}}. 
The measurements of the spectrum of emitted photons by hot naphthalene molecules 
in Ref. \cite{Saito2020} show a spectrum which is significantly broadened compared 
to the ground-state absorption case. 
This experimental observation agrees qualitatively with our calculations shown in 
Fig. \ref{Spectrum}. 

Measurements of thermal emission spectra of highly excited clusters provide a method 
to explore the potential energy surfaces of molecules and cluster, which is presently 
not available with other methods, such as traditional spectroscopy.
The results calculated here suggest that such measurements will give significant effects 
compared to ground state geometry spectra and emission rate constants. 
The ultimate information will be provided by experiments. 
The results here tell us that these experiments are worth performing.}
 
\section{Acknowledgements}
This work is supported by the Research Foundation Flanders 
(FWO, G.0A05.19N){\color{blue}, and the KU Leuven Research Council 
(project C14/22/103)}. 
PF acknowledges the FWO for a senior postdoctoral grant.
KH acknowledges support from the National Science Foundation of China with the 
grant 'NSFC No. 12047501' and the Ministry of Science and Technology of People's 
Republic of China with the 111 Project under Grant No. B20063. 
{\color{blue}T. H. is grateful for the J{\'a}nos Bolyai Research Scholarship of the Hungarian 
Academy of Sciences (grant number BO/00642/21/7).}
\bibliographystyle{apsrev4-2}
%
\end{document}